\documentclass[amsfonts,aps,nofootinbib,preprint]{revtex4}

\begin{document}

\title{Entanglement and entropy operator for
strings in pp-wave time dependent background}

\author{{A. L. Gadelha\footnote{gadelha@ift.unesp.br}},
D\'afni Z. Marchioro{\footnote{dafni@ift.unesp.br}} and Daniel
L. Nedel{\footnote{daniel@ift.unesp.br}}}

\affiliation{Instituto de F\'{\i}sica Te\'{o}rica,
UNESP - S\~ao Paulo State University, Rua Pamplona
145, S\~{a}o Paulo, SP, 01405-900, Brazil }

\begin{abstract}
In this letter new aspects of string theory
propagating in a pp-wave time dependent background
with a null singularity are explored. It is shown
the appearance of a 2d entanglement entropy
dynamically generated by the background.
For asymptotically flat observers, the vacuum
close to the singularity is unitarily inequivalent to the vacuum at
$\tau = -\infty$ and it is shown that the 2d entanglement
entropy diverges close to this point.
As a consequence, the positive time region is inaccessible for
observers in $\tau =-\infty$. For a stationary measure, the vacuum at
finite time is seen by those observers as a thermal state and the
information loss is encoded as a heat bath of string states.

\end{abstract}

\maketitle

\section{Introduction}
One of the most interesting subjects in string theory is its study
in time dependent backgrounds. Such endeavour may provide answers to
important questions involving typical quantum cosmology problems from
the string theory point of view, such as the initial conditions,
the cosmological singularities and the pre-big bang scenario.

In general, string theory in time dependent backgrounds is hard and
demands further development of its non perturbative sector. On the
other hand, string theory in time dependent plane wave backgrounds
is always in the perturbative regime and it is exactly solvable,
providing insights in many aspects of the theory
in more complicated geometries
\footnote{For a historical development of the subject see, for example, \cite{amati, horo, brooks, sanchez}.}.
In particular, in \cite{tsey} the string
theory is quantized in a pure dilatonic singular plane wave metric, where the worldsheet model exhibits many interesting
properties (or troubles) of field theory in curved space, including
the choice of the vacuum and particle creation (in this case string
mode creation). In the Brinkmann coordinates the metric is asymptotically flat
and defined in the interval $- \infty\leq \tau\leq +\infty$ of
the light-cone time, with a null singularity at the origin
($\tau =0$). So the model shares many characteristics with the
Gasperini and Veneziano's pre-big bang cosmology \cite{gaspvene},
and it can be used as a toy model to investigate this
scenario.\footnote{See also the Maurizio Gasperini's home page
http://www.ba.infn.it/~gasperin/.}

In this work we analyze the results of \cite{tsey} in Brinkmann
coordinates to explore different aspects of time dependent
backgrounds in the worldsheet context: the study of entanglement entropy
and loss of information. Concerning closed string propagating
in a pp-wave time dependent background, with asymptotically
flat regions ($\tau = \pm\infty$), it was shown in \cite{tsey}
(and references therein) that there is creation of string modes,
when the string propagates from one of the asymptotically flat
regions to the singularity in $\tau = 0$, as seen by an observer
in the flat vacuum $|0(\tau= \pm\infty), p_v\rangle$. This result
is of course observer dependent. In this letter we show
that the interaction of the string with the time dependent
background, as seen by a flat observer, creates a state with
a particular structure of excitation, characterized by a
condensation of right and left moving string modes.
This particular structure in a finite $\tau$ is the core of
the development we intend to do, presenting
a scenario where the entanglement generated by the background, the
entanglement entropy, loss of information and macroscopic
effects (thermal) can be studied.

The establishment that the vacuum state at
finite $\tau$ is a condensed state is possible due to the
observation that there is a Bogoliubov transformation mapping
the asymptotically flat string Hilbert space to its
corresponding one at a finite $\tau$. This Bogoliubov
transformation is formally identical to the general unitary
SU(1,1) Thermo Field Dynamic (TFD) transformation, proposed and analyzed in
\cite{su11,su11e,ion}\footnote{Recently in \cite{posg}
a general unitary $SU(2)$ TFD formulation was also
proposed to study type IIB Green-Schwarz superstring
at finite temperature.}. However, this identification
is just formal.
In TFD, there is a duplication of the system's degrees of
freedom by means of introducing an auxiliary system
(copy of the original (physical) one), followed by a
Bogoliubov transformation that entangles elements of the
two subspaces (original and auxiliary)\footnote{ Recently
it was shown that TFD states, seen as pure states, are
maximum entangled states \cite{admes}.}. In the case we are
studying, there is no duplication of the system's degrees
of freedom and the Bogoliubov transformation entangles the
left and right moving string modes in such a way that our
condensed state is in fact an entangled state of these modes.

It is known that in TFD the condensed state can be generated
by an entropy operator. This operator can be interpreted as an
entanglement entropy operator and has been used in different physical
situations, where TFD-like condensed states appear
\cite{garvit, fevit, ceravi, double, iorio}.
As we also have a condensed state with a TFD structure, we can look
for an entropy operator which generates this state and explore the
physical consequences of this fact. Actually we show that the entropy 
operator appears in a different context: it represents a 
2d entanglement entropy, dynamically generated by the time dependent 
background. This comes from the fact that,
although this time dependent geometry does not produce string
creation, from the worldsheet point of view, the metric produces
creation of string modes and all the information about the background
is contained in the entropy operator.  

We analyze the entropy operator expectation value at the
singularity and we show that it diverges owing to the infinite
creation of string modes. Even when the
singularity is avoided by an analytic continuation,
the behaviour remains the same. Furthermore, we show that the string's
Hilbert space close to the singularity is non unitarily related
to the flat Hilbert space, defined in $\tau=-\infty$. This is
also a result of the infinite creation of string modes close
to the singularity. In summary, although the singularity can
be avoided by an analytical continuation of the Bessel
functions, for observers defined in $\tau=-\infty$, there is
a complete loss of information about the string states at
the singularity and the positive time region is inaccessible.
 
Finally, we look for macroscopic effects related to the
generation of entropy and the respective information loss.
We show that in a particular (stationary) situation
the expectation value of the flat number operator in
the vacuum at finite time acquires the form of a Boltzmann
distribution by an appropriate choice of parameters. This
result is interpreted as a heat bath perceived by the
asymptotically flat observer when the string evolves in
the pp-wave time-dependent background.

The letter is divided as follows: in section two, we review
the model and fix the notation; in sections three and four we present
our results, and section five is devoted to conclusion and
discussion of the results.
 
\section{The Model}

In this section, we present a review of the results
obtained in \cite{tsey} for the model we want to
analyze\footnote{The time dependent pp-wave background
was also recently studied in the presence of a
Ramond-Ramond 5-form field strength \cite{chinas}. See also \cite{sumit1, sumit2}.}.
We are going to consider type IIB Green-Schwarz
superstring in the following  pp-wave time-dependent
background (in Brinkmann coordinates)
\begin{eqnarray}
ds^2 = 2dudv -\lambda x^2du^2 + dx^idx^i,
\nonumber
\\
\phi = \phi(u), \qquad \lambda = \frac{k}{u^2},
\end{eqnarray}
where $u$, $v$ are the light-cone coordinates, $x^i$ ($i=1\ldots d$)
are the transverse coordinates and $\phi$ is the dilaton. This
metric has a null singularity at $u=0$. Choosing the light-cone
gauge to fix the reparametrization invariance,
\begin{equation}
u = 2\alpha^{'}p^u \tau \:,
\end{equation}
the bosonic part of the action is
\begin{equation}
S = -\frac{1}{4\pi\alpha^{'}}
\int d\tau \int^{\pi}_{0}d\sigma
\left(\partial^{a}x^i\partial_{a}x^{j}\delta_{ij} +
\frac{k}{\tau^2}(x^i)^2 \right) \:,
\end{equation}
and the equations of motion for the transverse
coordinates read
\begin{equation}
(\partial^{2}_{\tau} - \partial^{2}_{\sigma})x^i +
\frac{k}{\tau^2}x^i = 0 \:.
\end{equation}

If we perform a change of coordinates from Brinkmann
to Rosen,
\begin{eqnarray}
v = V + \frac{1}{2}h(u)X^iX^i \:, \:\:\: x^i = a(u)X^i
\:, \nonumber\\
h = -a\frac{da(u)}{du} \:, \:\:\: \frac{d^2a(u)}{du^2}
= -\lambda(u)a(u) \:,
\end{eqnarray}
the metric takes the form
\begin{eqnarray}
ds^2 = 2dudV + a^2(u)dX^idX^i \:,\nonumber\\
a(u)= u^{1-\nu} \:, \:\:\: \nu =
\frac{1}{2}(1 + \sqrt{1-4k}) \:,
\:\:\: 0<k<\frac{1}{4}
\end{eqnarray}
and the equations of motion for the transverse
coordinates $X^i$ resemble the equation of a damped
harmonic oscillator \cite{tsey},
\begin{equation}
\left(\partial^{2}_{\tau} + \frac{\gamma}{\tau}\partial_{\tau}
- \partial^{2}_{\sigma}\right)X^{i} = 0 \:, \:\:\: \gamma = 2(1-\nu)
\end{equation}
with friction coefficient proportional
to $\frac{\gamma}{\tau_0}$ in a small
region of time around some $\tau_0$. So,
in Rosen coordinates it is clear that this metric produces
dissipation in the 2d worldsheet field theory.
As the entropy production and loss of information
appear naturally in some dissipative systems
\cite{ceravi,double}, the Rosen coordinate system
may be the natural one to make the analysis we
intend to do. However, we are going to work in
Brinkmann coordinates because it
is asymptotically flat, which makes the analysis easier.

The solution of the equations of motion in Brinkmann coordinates can
be written as
\begin{eqnarray}
x^i(\sigma , \tau) &=& x^i_0(\tau) + \frac{1}{2} \sqrt{2\alpha^{'}}
\sum^{\infty}_{n=1}\frac{1}{\sqrt{n}}
[Z(2n\tau)(a^{i}_{n}e^{2in\sigma} + \tilde{a}^{i}_{n}e^{-2in\sigma})
\nonumber
\\
&+& Z^{\ast}(2n\tau)(a^{i\:\dagger}_{n}e^{-2in\sigma} +
\tilde{a}^{i\:\dagger}_{n}e^{2in\sigma})]
\end{eqnarray}
with
\begin{eqnarray}
Z(2n\tau) \equiv e^{-i\frac{\pi}{2}\nu} \sqrt{\pi n
\tau}[J_{\nu-\frac{1}{2}}(2n\tau) -iY_{\nu-\frac{1}{2}}(2n\tau)]
\label{Zdef}
\\
x^{i}_{0}(\tau) = \frac{1}{\sqrt{2\nu -1}}(\tilde{x}^{i}\tau^{1-\nu}
+ 2\alpha^{'}\tilde{p}^{\: i}\tau^{\nu})
\\
\tilde{x}^{i} = \sqrt{\frac{\alpha^{'}}{2}}(a^{i}_{0} +
a^{i\dagger}_{0}), \qquad \tilde{p}^{\: i} =
\frac{1}{i\sqrt{2\alpha^{'}}}(a^{i}_{0} - a^{i\dagger}_{0})\:,
\end{eqnarray}
where $J_{\nu-\frac{1}{2}}$ and $Y_{\nu-\frac{1}{2}}$ are the Bessel
functions of the first and second type, respectively. The
creation/annihilation operators satisfy the usual oscillator algebra
\begin{equation}
[a^{i}_{n},a^{j\:\dagger}_{m}] = [ \tilde{a}^{i}_{n},
\tilde{a}^{j\:\dagger}_{m}] = \delta^{ij}\delta_{nm}, \qquad
[a^{i}_{0},a^{j\:\dagger}_{0}] = \delta^{ij},
\end{equation}
and all the other commutation relations vanish.

The Hamiltonian for the system under consideration can be written as
\begin{eqnarray}
H = \frac{1}{\alpha^{'}p_{v}}\left({\cal H}_{0}(\tau)
+\frac{1}{2}\sum^{\infty}_{n=1}n
[\Omega_{n}(\tau)(a^{\dagger}_{n}\cdot a_{n} +
\tilde{a}^{\dagger}_{n}\cdot\tilde{a}_{n} + D) -
B_{n}(\tau)a_{n}\cdot\tilde{a}_{n} -
B^{\ast}_{n}(\tau)a^{\dagger}_{n}\cdot\tilde{a}^{\dagger}_{n}]\right)
\nonumber
\\
\Omega_{n}(\tau) = \left(1 + \frac{\nu}{4\tau^{2}n^{2}}\right)\mid Z\mid^{2} +
\mid W\mid^{2} - \frac{\nu}{2n\tau}(ZW^{\ast} + Z^{\ast}W),
\nonumber
\\
B_{n}(\tau) = \left(1 + \frac{\nu}{4\tau^{2}n^{2}}\right)Z^{2} + W^{2}
-\frac{\nu}{\tau n}Z W,
\label{h0}
\end{eqnarray}
where $D=\delta^{ii}$. The dot denotes the scalar product
in the transverse space and
\begin{equation}
W(2n\tau) \equiv e^{-i\frac{\pi}{2}\nu} \sqrt{\pi n\tau}[J_{\nu +
\frac{1}{2}}(2n\tau) -iY_{\nu + \frac{1}{2}}(2n\tau)]. \label{Wdef}
\end{equation}
The term ${\cal H}_{0}$ in the first line of (\ref{h0})
is the zero-mode part, which is obtained treating the zero mode as a
harmonic oscillator with time-dependent frequency \cite{tsey}:
\begin{equation}
{\cal H}_{0}(\tau) = \frac{\alpha^{'}}{2}\left[(p^{i}_{0})^{2} +
\frac{k}{4\alpha^{'2}\tau^{2}}(x^{i}_{0})^{2}\right].
\end{equation}

The Hamiltonian, as expressed in terms of the modes $a_{n}$,
$\tilde{a}_{n}$, $a_0$, is non-diagonal. However, in the limit
$\tau = -\infty$ the Hamiltonian is diagonal and the metric is flat. 
In order to diagonalize it in a finite time, a new set of
time-dependent string modes is defined as
\begin{eqnarray}
{\cal A}^{i}_{n}(\tau) = f_{n}(\tau)a^{i}_{n} +
g^{\ast}_{n}(\tau)\tilde{a}^{i\:\dagger}_{n}, \qquad {\cal
A}^{i\dagger}_{n}(\tau) = f^{\ast}_{n}(\tau)a^{i\:\dagger}_{n} +
g_{n}(\tau)\tilde{a}^{i}_{n}, \nonumber
\\
\tilde{\cal A}^{i}_{n}(\tau) = f_{n}(\tau)\tilde{a}^{i}_{n} +
g^{\ast}_{n}(\tau)a^{i\:\dagger}_{n}, \qquad \tilde{\cal
A}^{i\dagger}_{n}(\tau) =
f^{\ast}_{n}(\tau)\tilde{a}^{i\:\dagger}_{n} + g_{n}(\tau)a^{i}_{n},
\label{acal}
\end{eqnarray}
where
\begin{eqnarray}
f_{n}(\tau) =
\frac{1}{2}\sqrt{\frac{w_n}{n}}e^{2iw_{n}\tau}
\left[Z(2n\tau) + \frac{i}{2w_{n}}\dot{Z}(2n\tau)\right]\:,
\nonumber
\\
g_{n}(\tau) = \frac{1}{2}\sqrt{\frac{w_n}{n}}e^{-2iw_{n}\tau}
\left[Z(2n\tau) - \frac{i}{2w_{n}}\dot{Z}(2n\tau)\right],
\label{fg}
\end{eqnarray}
and
\begin{eqnarray}
w_{n}(\tau) = \sqrt{n^{2} + \frac{k}{4\tau^{2}}}\:, \nonumber\\
\dot{Z}(2n\tau)=\partial_{\tau}Z(2n\tau) = \frac{\nu}{\tau}Z(2n\tau)
- 2nW(2n\tau)\:,
\label{zdot}
\end{eqnarray}
with $Z(2n\tau)$ and $W(2n\tau)$ defined in (\ref{Zdef}) and
(\ref{Wdef}), respectively.

We can see that this diagonalization process is in fact a canonical
transformation, since
\begin{equation}
\left|f_{n}(\tau)\right|^2 - \left|g_{n}(\tau)\right|^2 =1,
\end{equation}
and the commutation relations for the new set of modes are
\begin{equation}
[{\cal A}^{i}_{n}(\tau), {\cal A}^{j\dagger}_{m}(\tau)] =
\delta_{nm}\delta^{ij} \:, \:\:\: [\tilde{\cal A}^{i}_{n}(\tau),
\tilde{\cal A}^{j\dagger}_{m}(\tau)] = \delta_{nm}\delta^{ij} \:,
\:\:\: [{\cal A}^{i}_{n}(\tau),
 \tilde{\cal A}^{j\dagger}_{m}(\tau)] = 0.
\end{equation}

Having established the model and the notation,
let's deal with the object of our paper.

\section{Dynamical Entanglement and Entropy Operator}

In this section we are going to analyze the model
presented above as seen by an observer in the
vacuum $|0,p_v \rangle$ at $\tau=-\infty$. 
The analysis consists of beginning in the vacuum
at $\tau = -\infty$ and studying how this state
evolves to $\tau =0$. To this end we are going to
construct a Bogoliubov transformation, relating the
Hilbert space in $\tau=-\infty$ with the Hilbert space
at a finite time. Notice that, owing to the symmetry of the metric
$\tau\rightarrow -\tau$, this system could also be
analyzed from $\tau =+\infty$ evolving back to $\tau =0$.
We consider the vacuum $\mid 0, p_v \rangle$  as the
flat vacuum annihilated by the $a_{n}^{i}$,
$\tilde{a}_{n}^{i}$ modes and look for the vacuum
$|0(\tau), p_v \rangle$ annihilated by
${\cal A}_{n}(\tau)$ and $\tilde{{\cal A}}_{n}(\tau)$.
We are going to show that this vacuum is a
coherent state defined by an entanglement entropy
operator.

The first step in our construction is to notice that
the expression (\ref{acal}) can be written as the
following general $SU(1,1)$ Bogoliubov transformation 
\begin{eqnarray}
\left(
\begin{array}{c}
{\cal A}^{i}_{n}(\tau) \\
\widetilde{{\cal A}}^{i \dagger}_{n}(\tau)
\end{array}
\right)
&=&{\mathbb B}_{n}(\tau)\left(
\begin{array}{c}
a^{i}_{n} \\
\widetilde{a}^{i \dagger }_{n}
\end{array}
\right) ,
\label{tbt}
\\
\left(
\begin{array}{cc}
{\cal A}^{i \dagger}_{n}(\tau) & -\widetilde{{\cal A}}^{i}_{n}(\tau)
\end{array}
\right) &=&\left(
\begin{array}{cc}
a^{i \dagger }_{n} & -\widetilde{a}^{i}_{n}
\end{array}
\right) {\mathbb B}^{-1}_{n}(\tau), \label{tbti}
\end{eqnarray}
where the $SU(1,1)$ matrix transformation is given by
\begin{eqnarray}
{\mathbb B}_{n}(\tau)=\left(
\begin{array}{cc}
f_{n}(\tau) & g^{\ast}_{n}(\tau) \\
g_{n}(\tau) & f^{*}_{n}(\tau)
\end{array}
\right) ,\qquad |f_{n}(\tau)|^{2}-|g_{n}(\tau)|^{2}=1,
\label{tbm}
\end{eqnarray}
in such a way that we have a canonical transformation mapping
asymptotically flat string operators into operators
defined in a finite $\tau$. Now, we are going to
find the transformation for the states, in order
to have a map between the flat Hilbert space and
the finite time one.
To this end it is interesting to recognize that
the structure of the Bogoliubov transformation
(\ref{tbt}), (\ref{tbti}) and (\ref{tbm}) is the same
as the general unitary $SU(1,1)$ (TFD) formulation
\cite{su11,su11e,ion}.
However, as mentioned in the introduction, here
there is no duplication of the degrees of
freedom by introducing an auxiliary non-physical system 
and both set of oscillators are dynamical one. 
We can take advantage of the well-known
SU(1,1) structure cited above to deal
with the system in a schematic way. For
example, using the polar decomposition
for the Bogoliubov matrix elements,
(\ref{tbm}) can be rewritten as
\begin{equation}
{\mathbb
B_{n}}(\tau)=\frac{e^{s_{n}\tau_{3}}}{\sqrt{1-\overline{f}_{n}}}\left(
\begin{array}{cc}
1 & -\overline{f}_{n}^{\alpha_{n}} \\
-\overline{f}_{n}^{1-\alpha_{n}} & 1
\end{array}
\right), \label{tbmr}
\end{equation}
where $\tau_{3}$ is a Pauli matrix,
\begin{eqnarray}
\overline{f}_{n}=\overline{f}_{n}(\tau)
=\frac{|g_{n}(\tau)|^{2}}{|f_{n}(\tau)|^{2}},
\qquad
\alpha_{n}=\alpha_{n}(\tau)
=\frac{\ln(-\frac{g^{\ast}_{n}(\tau)}{f_{n}(\tau)})}{\ln(\overline{f}_{n}(\tau))}
=
\frac{1}{2}-i\frac{(\phi_{n}(\tau)+\varphi_{n}(\tau))}{\ln(\overline{f}_{n}(\tau))}, \nonumber\\
s_{n}=s_{n}(\tau)=i\varphi_{n}(\tau)
=\frac{1}{2}\ln\left(\frac{f_{n}(\tau)}{f_{n}^{*}(\tau)}\right), \label{sfap}
\end{eqnarray}
and $\alpha_{n} + \alpha^{\ast}_{n}=1$. Furthermore we can write
down a state annihilated by ${\cal A}^{i}_{n}(\tau)$ and
$\tilde{\cal A}^{i}_{n}(\tau)$ at each $\tau$, and interpret it as
the vacuum of the Bogoliubov transformed system.
Such a state has the structure of a condensed state of
$a^{i \dagger}_{n}\widetilde{a}^{i \dagger }_{n}$-pair and,
by construction, it is also a superposition of
$SU(1,1)$ coherent states
\begin{equation}
|0(\tau), p_v \rangle = \prod_{n=1}^{\infty}
(1-\overline{f}_{n})^{\alpha_n D}
e^{\overline{f}^{\,\,\alpha_n }_{n} a_{n}^{\dagger}\cdot
\widetilde{a}_{n}^{\dagger}} \mid 0,p_v\rangle,
\label{ets}
\end{equation}
with its dual state being
\begin{equation}
\langle 0(\tau), p_v |= \langle 0,p_v\mid
\prod_{n=1}^{\infty} (1-\overline{f}_{n})^{(1-\alpha_n ) D}
e^{\overline{f}^{\,\,1-\alpha_n }_{n}a_{n}\cdot\widetilde{a}_{n}} \:.
\end{equation}
Since
$\overline{f}_{n}^{\;\alpha_n}(\tau)
=-\frac{g^{\ast}_{n}(\tau)}{f_{n}(\tau)}$, one can see that
\begin{equation}
{\cal A}^{i}_{n}(\tau)|0(\tau),p_v \rangle=
\tilde{\cal A}^{i}_{n}(\tau)|0(\tau), p_v \rangle =0,
\end{equation}
and in fact the condensed state (\ref{ets}) is the vacuum
of the system at a finite time. 
The structure of the state (\ref{ets}) shows us that,
owing to this time dependent background, the
string is excited in a particular way characterized by
a condensed state of left and right moving string modes. Actually, this 
condensed state is an entangled
state of string modes similar to TFD states. 
However,  it is important to stress the crucial difference between
the usual TFD and the use of its structure presented here: 
while in TFD the entanglement is between
the original system with an auxiliary one \cite{UME},
here the Bogoliubov transformation entangles the
right and left moving string oscillators.
Besides, the string entangled state is generated 
by the background and it is observer dependent.

Again inspired in the TFD, it is possible to show that
the entangled state defined in (\ref{ets}) can be
obtained from the following expression:
\begin{equation}
|0(\tau), p_v \rangle = e^{i\Psi(\tau)}e^{-\frac{\bf{K}}{2}}
e^{\sum_{n}a_{n}^{\dagger}\cdot \widetilde{a}_{n}^{\dagger}} \mid
0,p_v\rangle, \label{vack}
\end{equation}
where $\Psi(\tau)$ is
\begin{equation}
\Psi(\tau)=
\sum_{n=1}^{\infty}
\left[\frac{(\phi_{n}(\tau)+\varphi_{n}(\tau))}
{\ln(\overline{f}_{n}(\tau))}{\mathbf K}_{n}\right]
\end{equation}
and $\bf{K}$ is
\begin{equation}
{\bf K} =\sum_{n=1}^{\infty}{\mathbf K}_{n}
=\sum_{n=1}^{\infty}\left( K_{n} + \widetilde{K}_{n}\right),
\end{equation}
with
\begin{equation}
K_{n} = -\left[a_{n}^{\dagger}\cdot {a}_{n}
\ln\left(|g_{n}(\tau)|^{2}\right)-a_{n}\cdot
{a}_{n}^{\dagger}\ln\left(|f_{n}(\tau)|^{2}\right)\right] \:.
\end{equation}
The expression for $\widetilde{K}_{n}$ is obtained changing
$a_{n}, a^{\dagger}_{n}$ for $\widetilde{a}_{n},
\widetilde{a}^{\dagger}_{n}$ in $K_{n}$. Now we have a operator mapping 
$|0,p_v\rangle$ into $|0(\tau),p_v\rangle$.
In the TFD approach, the $\bf{K}$ is called
entropy operator since its expectation value in the
thermal vacuum furnishes the thermodynamical entropy
of the system. Here it can be viewed as the generator of the
entangled state (\ref{ets}) and it is called in this
context the entanglement entropy operator
\cite{ceravi,iorio}. Nevertheless, the structure of the string
entanglement generated by the entropy operator differs from the one
used in \cite{ceravi,iorio} by complex parameters. In fact, we have a
more general structure since the entanglement of \cite{ceravi,iorio}
is related to the usual TFD formulation, while in this work the entropy 
operator is related to the general unitary SU(1,1) formulation of TFD. 
As it is shown in \cite{umeb2,hen}, the expectation values
do not depend on $\alpha_n$ and we do not need to worry about the
imaginary part of the exponentials defined in (\ref{vack}).
For example, if we consider the ${\bf K}$ expectation value in
$|0(\tau), p_v \rangle$, we find
\begin{equation}
{\cal S}(\tau) = -2d\sum_{n=1}^{\infty}[|g_{n}(\tau)|^2\ln ( |g_{n}(\tau)|^2)
-(1+|g_{n}(\tau)|^2 )\ln (1 +|g_{n}(\tau)|^2 )] \:.
\label{s}
\end{equation}
since $\langle 0(\tau),p_v | 0(\tau), p_v \rangle =1$ and
\begin{equation}
\langle 0(\tau), p_v | N_{n} |0(\tau),p_v \rangle =\langle 0(\tau),p_v |
\widetilde{N}_{n} |0(\tau),p_v \rangle = D |g_{n}(\tau)|^{2}.
\end{equation}
for $N_{n}=a^{\dagger}_{n} \cdot a_{n}$ and
$\widetilde{N}_{n}=\widetilde{a}^{\dagger}_{n} \cdot
\widetilde{a}_{n}$.

Noticing that (\ref{vack}) can have the alternative form
\begin{equation}
|0(\tau),p_v \rangle= e^{i\Psi(\tau)}\sum_{n_k
=0}^{\infty}\prod_{k}
\left(\frac{|g_{k}(\tau)|^{2n_{k}}}{|f_{k}(\tau)|^{2n_{k}+2}}\right)^{D}\mid
n_k ,\widetilde{n}_k ,p_{v}\rangle , \label{kvac2}
\end{equation}
the expectation value of ${\bf K}$ can also be written as
\begin{equation}
{\cal S}(\tau)= \sum_{n}{\cal W}_{n}(\tau) \ln {\cal W}_{n}(\tau), 
\end{equation}
where
\begin{equation}
{\cal W}_{n}(\tau)=\prod_{k}
\left(\frac{|g_{k}(\tau)|^{2n_{k}}}{|f_{k}(\tau)|^{2n_{k}+2}}\right)^{D},
\label{wn}
\end{equation}
are the eigenvalues of the system's density matrix.

In order to investigate the entropy's behaviour as $n|\tau| \rightarrow
0$ and $n|\tau| \rightarrow \infty$, note that $|g_{n}(\tau)|^2$ can be
expressed in terms of $\Omega_n$ and $w_n$ given in (\ref{h0})
and (\ref{zdot}) respectively:
\begin{equation}
|g_n (\tau)|^{2} = \frac{1}{4w_n}(n\Omega_{n} - 2w_n)\:.
\end{equation}
So, it is just needed to know how $\Omega_n$ behaves in these two
limits. For $n|\tau| \gg 1$, one has \cite{tsey}
\begin{equation}
\Omega_{n}(\tau) = 2 + \frac{k}{4n^2 \tau^2} - \frac{k^2}{64n^4
\tau^4} + \frac{k^2 (2+k)}{512n^6 \tau^6} + \ldots \:,
\end{equation}
and, after some cancellations, $|g_n (\tau)|^2$ takes the form
\begin{equation}
\mid g_n (\tau)\mid^2 \:\cong \frac{k^2}{1024\tau^6 n^6} \:.
\end{equation}

\noindent Substituting the expression above in (\ref{s}), the
dependence of the entropy with respect to $\tau$ is
\begin{equation}
{\cal S}(\tau) \approx \frac{D\:k^2}{483840\tau^6}\times \left[\pi^6
\ln\left(\frac{1024\tau^6}{k^2}\right) - 5670\zeta^{'}(6)\right] \:,
\end{equation}

\noindent where $\zeta^{'}(6)$ is the first derivative of the Zeta
function. The result obtained above guarantees that the entropy is
positive definite and it increases when goes from $\tau = -\infty$ to
0. If we take the limit $|\tau| \rightarrow \infty$, the entropy tends to
zero, as expected for the asymptotically flat states.

Now we analyze ${\cal S}$ close to the singularity ($n|\tau| \ll 1$). In
this case, $\Omega_n$ takes the form
\begin{equation}
\Omega_n (\tau) \cong \frac{\pi}{(n\tau)^{2\nu}\cos^2
(\pi\nu)}\left(\frac{1}{[\Gamma(\frac{1}{2}-\nu)]^2} +
\frac{\nu}{4[\Gamma(\frac{3}{2}-\nu)]^2} +
\frac{\nu}{\Gamma(\frac{3}{2}-\nu)\Gamma(\frac{1}{2}-\nu)}\right)\:.
\label{omega}
\end{equation}

\noindent or
\begin{equation}
\Omega_n \approx C \times \frac{1}{(n\tau)^{2\nu}}
\end{equation}

\noindent where $C$ is the constant part of (\ref{omega}). As $n|\tau|
\ll 1$, $w_n \rightarrow \frac{\sqrt{k}}{2\mid \tau\mid}$, we have:

\begin{eqnarray}
{\cal S}(\tau) =
-2D\sum_{n=1}^{\infty}
\left[\left(\frac{C}{2\sqrt{k}}\times(n\mid \tau\mid)^{1-2\nu}
- \frac{1}{2}\right)
\ln \left(\frac{\frac{C}{2\sqrt{k}}\times(n\mid \tau\mid)^{1-2\nu}
- \frac{1}{2}}{\frac{C}{2\sqrt{k}}\times(n\mid \tau\mid)^{1-2\nu}
+ \frac{1}{2}}\right)\right. \nonumber\\
\left.-\ln \left(\frac{C}{2\sqrt{k}}\times(n\mid \tau\mid)^{1-2\nu}
+
\frac{1}{2}\right)\right] \:. \label{s1}
\end{eqnarray}

\noindent One can see from the equation above that the entropy diverges
when $n|\tau| \ll 1$. This behaviour of ${\cal S}(\tau)$ close to the
singularity is owing to the creation of high frequency modes.
Note that it is possible to avoid the singularity using an
analytic continuation of the Bessel functions, although this
procedure leads to a discontinuity in the time derivative in the
zero mode sector \cite{tsey}. However, using this approach,
the entropy operator does not change and
therefore it still diverges. This means that, for
an observer in $\tau=-\infty$, there is a complete
loss of information when the string approaches the
singularity, and the positive time region turns out to be
inaccessible. This is clear if we calculate the following projection
\begin{equation}
\langle p_{v}, 0 \mid 0(\tau),p_v \rangle= e^{-D
\sum_{n}\ln\left(1+|g_{n}(\tau)|^{2}\right)}e^{2iD\sum_n\left[
\left(\phi_{n}(\tau)+\varphi_{n}(\tau)\right)\frac{\ln|f_n(\tau)|^2}{\ln(\overline{f}_{n}(\tau))}
\right]} \label{ket}
\end{equation}

\noindent and take the limit $n|\tau| \ll 1$ . The result is $\langle
p_{v}, 0 \mid 0(\tau),p_v \rangle = 0$, which shows that in this limit the
states at different times are unitarily inequivalent to the vacuum
at $\tau = -\infty$. In other words, this result shows that close
to the singularity, the system defined by the 2d worldsheet quantum
field theory is led to another representation of the canonical
commutation relations, which is unitarily inequivalent to the
representation at $\tau =-\infty$. We will return to this point
in the conclusions. 

\section{Thermal effects}

We have shown that an entanglement entropy
appears when the string propagates in this
time-dependent geometry. The loss of information related
to such a behaviour can be manifested macroscopically and
it is natural to search for thermal effects in this context.
It can be achieved if we take advantage of the
$SU(1,1)$ structure of the system and identify
\begin{equation}
\sinh^2 (\theta_n)
= |g_n (\tau)|^2 \:, \qquad \cosh^2 (\theta_n) = |f_n (\tau)|^2 \:,
\end{equation}
where the parameters $\theta_n$ encode the information of
all the relevant parameters of the theory.
Let's define the following potential as a Legendre transformation
of the entanglement entropy
\begin{equation}
{\cal F}(\beta,T)={\cal E(\tau)} - T {\cal S(\tau)},
\end{equation}
where
${\cal E}=
\langle 0(\tau),p_v |H_{flat}|0(\tau),p_v \rangle$ 
is the expectation value of the asymptotically flat ($\tau = -\infty$)
Hamiltonian operator, and
$T =\partial {\cal E}/\partial{\cal S}$
is a positive definite parameter. If we look at values of
$\theta_n$ that make ${\cal F}$ stationary and considering negligible
the variations of $T$ with respect to time, we have
\begin{equation}
|g_n (\tau)|^2 =
\sinh^{2}(\theta_{n})
=\frac{1}{e^{\frac{1}{T}\left(\frac{n}{\alpha^{'}p_v}\right)}-1}.
\label{bd}
\end{equation}
For this configuration of $|g_n (\tau)|^2$ the vacuum
$|0(\tau),p_v \rangle$ is a maximal entangled state
as can be proved following Ref. \cite{admes}.
Under such a specific configuration that maximizes
the entanglement entropy, the expression (\ref{bd})
can be viewed as the Boltzmann distribution for string modes
providing that $T$ is identified with the temperature.
This scenario can be interpreted as follows:
the loss of information's effect, as seen by an
asymptotically flat observer, is macroscopically
manifested as a heat bath when the string evolves
in the time-dependent background.

\section{Conclusions}

In this work the closed string propagating in a pure dilatonic time
dependent background is studied in a new perspective, where the
production of entangled states is explored.
In Brinkmann coordinates, the evolution of the string
states from $\tau=-\infty$ to the singularity at $\tau=0$ is analyzed.
From the point of view of
an observer in the flat vacuum, the
vacuum in a finite time is a $SU(1,1)$ entangled state of right and
left moving string modes defined by an entropy operator,
which plays the role of the entanglement generator. 
Its expectation value, evaluated in the finite time vacuum, furnishes 
the entanglement entropy perceived by the flat observer. 
In addition, it was shown that close to the singularity the
entanglement entropy seen by the observer in flat vacuum diverges. 
Also, we showed that the Hilbert space close to the singularity is 
unitarily inequivalent to the flat Hilbert space.  
These results imply a complete loss of information at the singularity 
and consequently the positive time region is inaccessible to an
observer in $\tau=-\infty$. 

The unitary nonequivalence between the Hilbert spaces
is a typical result of dissipative motion, where the entropy operator
drives the system through unitarily inequivalent states and it is interpreted 
as a time evolution controller, as shown in \cite{ceravi,double}. At this
point it is interesting to remember that in Rosen coordinates the
string's equation of motion resembles a damped harmonic oscillator
equation. So, it is not surprising to find a dissipative behaviour in other 
coordinate system. Consequently, it is tempting to conjecture that 
the string entropy operator generates the time evolution of the system.  
However, there is a crucial difference between the results 
of \cite{ceravi,double} and our work. While in \cite{ceravi,double}
the condensed state generated by the entropy operator is 
also generated by a SU(1,1) Hamiltonian, in our work the string
condensed state 
is generated by a Bogoliubov transformation and we only have a
map between Hilbert spaces, 
which turns out to be non unitary close to the singularity.
Although the string Hamiltonian (\ref{h0}) also has a SU(1,1)
structure, we are not able, at present time, to conclude that the 
string entropy operator generates the evolution of the system.
This issue is now under investigation.

At the end it is shown how the loss of information can be
macroscopically manifested as a heat bath. By means of a
Legendre transformation of the entropy operator, we find
a particular stationary configuration which maximizes the
entanglement. Besides, this configuration allows us to
introduce a parameter that can be interpreted as the
temperature, and the expectation value of the number
operator at finite time acquires the form of a Boltzmann
distribution. So, in this configuration, for asymptotically
flat observers, all the information of the string states
in a finite time is encoded as a heat bath.

There are many possible extensions of this work. As there
is a complete loss of information at the singularity, the
approach used in this letter could provide tools to explore
the pre-big bang phase scenario from a different perspective,
where the pre-big-bang states may appear as a heat bath.
On the other hand, it will be interesting to see how the
string entropy operator used here can appear in the context
of reference \cite{brustein}, where it was given an
entanglement interpretation for black hole entropy in string
theory, using dual field theories. An entropy operator
was also used to study information loss in
classical dissipative system, in connection with the t'Hooft
deterministic quantum mechanics \cite{bla, thooft}.
We can further develop their techniques to investigate this
point in the string model studied here, since we have shown
that a 2d string entropy operator appears naturally owing
to time dependent geometry. 

\begin{acknowledgments}
The authors would like to thank Marcelo Botta Cantcheff
and  M. C. B. Abdalla for useful discussions and FAPESP
(Funda\c{c}\~ao de Amparo \`a Pesquisa do Estado de S\~ao Paulo)
for financial support. This paper is dedicated to Marina Gadelha.
\end{acknowledgments}

\end{document}